\documentclass[twocolumn]{aastex62}

\graphicspath{{./}{figures/}}
\newcommand*{\add}[1]{\textcolor{black}{#1}}
\newcommand*{\addd}[1]{\textcolor{black}{#1}}
\usepackage{enumerate}
\usepackage{enumitem}

\shortauthors{Lee et al.}

\begin{document}

\title{FIRST J1419+3940 as the First Observed Radio Flare from a Neutron Star Merger}

\author{K.H. Lee}
\affiliation{Department of Physics, University of Florida, 2001 Museum Road, Gainesville, FL 32611, USA}

\author{I. Bartos}
\thanks{imrebartos@ufl.edu}
\affiliation{Department of Physics, University of Florida, 2001 Museum Road, Gainesville, FL 32611, USA}

\author{G.C. Privon}
\affiliation{National Radio Astronomy Observatory, 520 Edgemont Road, Charlottesville, Va, 22903}

\author{J.C. Rose}
\affiliation{Department of Astronomy, University of Florida, 211 Bryant Space Sciences Center, Gainesville, FL
32611, USA}

\author{P. Torrey}
\affiliation{Department of Astronomy, University of Florida, 211 Bryant Space Sciences Center, Gainesville, FL
32611, USA}

\begin{abstract}
During their violent merger, two neutron stars can shed a few percent of their mass. As this ejecta expands, it collides with the surrounding interstellar gas, producing a slowly-fading {\it radio flare} that lasts for years. Radio flares uniquely probe the neutron star merger populations as many events from past decades could still be detectable. Nonetheless, no radio flare observation has been reported to date. Here we show that the radio transient FIRST J1419+3940, first observed in 1993 and still detectable, \add{could have} originated from a neutron star merger. We carry out numerical simulations of neutron star merger ejecta to demonstrate that the observed radio light curve is well reproduced by a merger model with astrophysically expected parameters. We \add{examine the observed radio data, as well as the host galaxy, to find clues that could differentiate the transient's neutron star merger origin} from the alternative explanation---the afterglow of an off-axis long gamma-ray burst. Near-future observations could \add{find further evidence} for the FIRST J1419+3940 radio transient's origin. We show that existing radio surveys likely already recorded multiple radio flares, informing us of the origin and properties of neutrons tar mergers and their role in the nucleosynthesis of the heaviest elements in the Universe.
\end{abstract}


\section{Introduction}

The multi-messenger discovery of neutron star merger GW170817 presented a remarkable lineup of emission processes expected from such mergers \citep{2017ApJ...848L..12A}. The gravitational wave signal showed that the neutron stars' masses and merger rate are consistent with expectations \citep{2017PhRvL.119p1101A}; gamma-ray observations confirmed that neutron star mergers produce short gamma-ray bursts (GRBs; \citealt{2017ApJ...848L..13A}); while the observed optical kilonova was produced by an ejecta with mass and velocity largely in line with theoretical predictions \citep{2017ApJ...848L..24V,2017Sci...358.1565E,2017Sci...358.1559K,2017ApJ...848L..17C}.

\begin{figure*}[]
\includegraphics[width=0.95\textwidth]{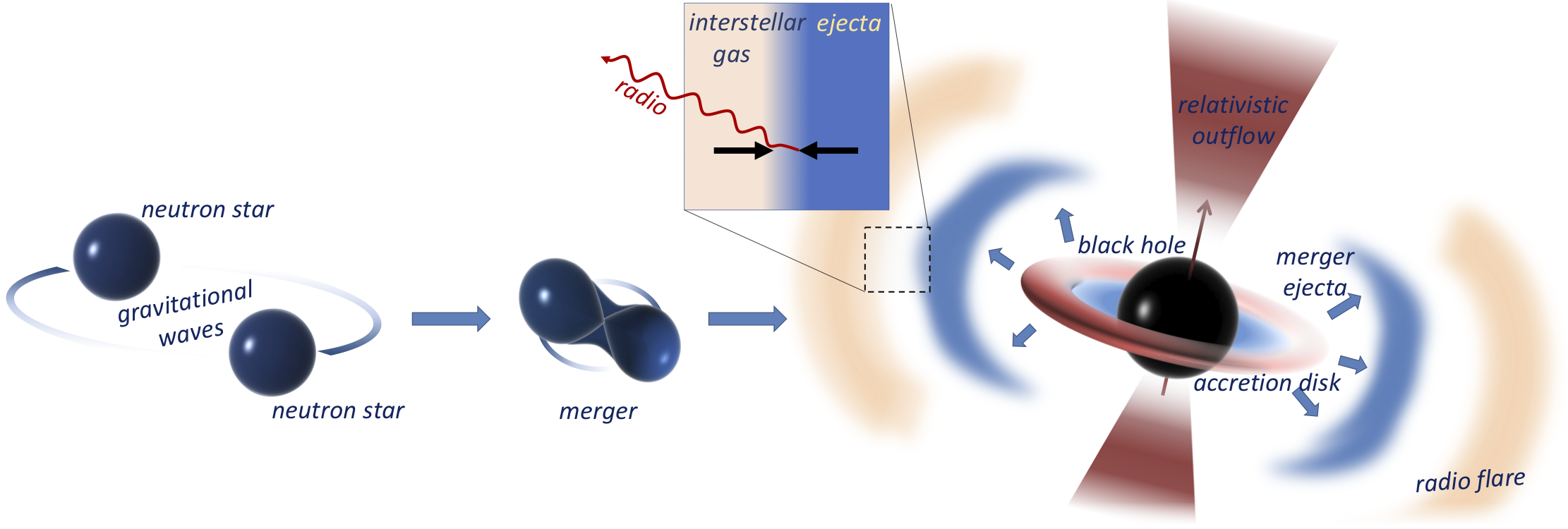}
\centering
\caption{Illustration of radio emission from a neutron star merger. During the merger, some neutron star matter gets ejected dynamically or through winds. This ejecta interacts with the interstellar gas, producing a years-long radio flare.}
\label{fig:illustration}
\end{figure*}

A notable exception has been a radio transient expected from the interaction of the kilonova ejecta with the surrounding interstellar medium (\citealt{2011Natur.478...82N}; see Fig. \ref{fig:illustration}). Such emission, which we will refer to as a {\it radio flare}, is emitted isotropically and can last for years after the merger, making it a promising target for follow-up observations \citep{2016ApJ...831..190H,2019MNRAS.485.4150B}. However, the radio flux strongly depends on the density of the interstellar medium, and detection is challenging at the low densities of $n\sim10^{-4}$\,cm$^{-3}$ found for GW170817 \citep{2019ApJ...870L..15L,2018ApJ...856L..18M}.

Radio flares from neutron star mergers thus remain an outstanding possibility yet to be discovered, not just for GW170817 but for any merger. A successful observation requires a nearby event within $\lesssim 200$\,Mpc \citep{2016ApJ...831..190H}, much closer than short GRBs identified so far other than GW170817 (albeit this could be due to selection effects \citep{2019MNRAS.485.4150B,2018arXiv180806238G}), and a circum-merger density of $\gtrsim10^{-2}$\,cm$^{-3}$, greater than the density for GW170817 and many other short GRBs \citep{2015ApJ...815..102F}.

FIRST J141918.9+394036 (hereafter J1419+3940) is a decades-long radio transient identified in a galaxy 87\,Mpc away from Earth \citep{law2018discovery}. It was first found by \cite{2017ApJ...846...44O} using the Very Large Array's (VLA) Faint Images of the Radio Sky at Twenty-Centimeters (FIRST) survey \citep{1995ApJ...450..559B}. Ofek searched for persistent radio-luminous sources in nearby galaxies in order to find long-term counterparts of fast radio bursts. The transient nature of J1419+3940 was later discovered by \cite{law2018discovery} using data from the NRAO-Very Large Array Sky Survey \citep{1998AJ....115.1693C}. Based on its radio light curve observed over 23 years, the high star formation rate of its host galaxy and the lack of a detected GRB counterpart, \cite{law2018discovery} explained the event as the afterglow of an off-axis long-duration GRB, i.e. one that is beamed away from Earth. They further suggested that the emission could be produced by a young magnetar. However, this latter possibility turned out to be inconsistent with the observed large source size \citep{2019ApJ...876L..14M}. 

In this paper we investigated whether the observed radio emission from J1419+3940 is consistent with a neutron star merger origin, and whether a merger origin is a more likely explanation for the transient than a GRB afterglow. In Section \ref{sec:methods} we show that the observed radio light curve can be well fit by a neutron star merger radio flare, and discuss consistency with the host galaxy. In Section \ref{sec:comparison} we discuss whether a neutron star merger radio flare and or a long GRB afterglow could be the better explanation of the observation. In Section \ref{sec:futureobs} we discuss the prospects of future radio observations in uncovering neutron star mergers. We conclude in Section \ref{sec:conclusion}.

\section{Neutron star merger model can explain observations} \label{sec:methods}

\subsection{Radio Observations}

We first probed whether J1419+3940 is consistent with a neutron star merger origin. We adopted the collected radio observations of \cite{law2018discovery} (see their Table 2.), and in addition the observation of \cite{marcote2019resolving} with the European VLBI Network. These observations have been carried out at a variety of radio frequencies. For the easier interpretation of the results, we converted the observed fluxes to their expected values at either 0.3\,GHz, 1.4\,GHz and 3\,GHz. All observations had a frequency close to one of these three values. Each result was converted to the closest of these three frequencies by assuming that the flux scales with frequency as $\nu^{-(p-1)/2}$, where $p$ is the power-law index of the distribution of the accelerated electrons' Lorentz
factors, which we take to be $p=2.5$ \citep{2013MNRAS.430.2121P}. The measured fluxes and upper limits as functions of time are shown in Fig. \ref{fig:radio}.

\subsection{Radio Flare Emission Model} \label{sec:emissionmodel}

We modeled the radio emission from the interaction of neutron star merger ejecta and the circum-merger medium following the prescription of \cite{2013MNRAS.430.2121P}. Based on the numerical outflow simulations of \cite{2017CQGra..34o4001F}, \add{we adopted a broken power-law velocity probability density for this outflow, with peak velocity $v_{0}\approx0.2$, and indices $\alpha_1=5$ and $\alpha_2=-10$ below and above the peak, respectively}. 

\add{We considered a uniform circum-merger medium with baryon number density $n$. We parametrized the expansion of the ejecta into this medium with the radial position $R$ of the front of the ejecta from the merger site. At a given $R$, the front of the outflow accumulated $4/3\pi R^3n+M_{\rm ej}(\beta)$ mass, where $M_{\rm ej}(\beta)$ is the part of the ejecta mass with initial velocity $\geq\beta$. Let $E(\beta)$ be the total kinetic energy of the part of the ejecta mass with initial velocity $\geq\beta$. Then the velocity $\beta=\beta(R)$ as a function of $R$ can be determined from energy conservation \citep{2013MNRAS.430.2121P}:}
\begin{equation}
M(R)(\beta c)^2 \approx E(\geq\beta).
\end{equation}
\add{We solved this equation numerically to obtain $\beta(R)$, and used this to compute $R(t)=\int_{0}^{t}\beta(R)^{-1}dR$, where time $t$ is measured from the start of the outflow.} 

\add{To determine the radio flux produced by the outflow, we computed two relevant characteristic frequencies following \cite{2013MNRAS.430.2121P}: the typical electron synchrotron frequency}
\begin{equation}
\nu_{\rm m}(t) \approx \textup{1\,GHz}\cdot n_{0}^{\frac{1}{2}}\epsilon_{\rm B, -1}^{\frac{1}{2}}\epsilon_{\rm e, -1}^{2}\beta(t)^{5}
\end{equation}
and the self-absorption frequency
\begin{equation}
\nu_{\rm a}(t) \approx \textup{1\,GHz}\cdot R_{17}(t)^{\frac{2}{p+4}} n_{0}^{\frac{6+p}{2(p+4)}}\epsilon_{\rm B, -1}^{\frac{2+p}{2(p+4)}}\epsilon_{\rm e, -1}^{\frac{2(p-1)}{p+4}}\beta(t) ^{\frac{5p-2}{p+4}}.
\end{equation}
Here, $\epsilon _{\rm B}$ and $\epsilon _{\rm e}$ are the fractions of the total internal energy of the shocked gas carried by the magnetic fields and electrons. In these equations and below we adopted the notation.

\add{The relation of these characteristic frequencies to the observational frequency $\nu_{\rm obs}$ affects the expected flux (see Table 2. in \citealt{2013MNRAS.430.2121P}). To compare with available observational data we computed the expected radio flux at 0.3\,GHz, 1.4\,GHz, and 3\,GHz (see Fig. \ref{fig:radio}). For frequencies 1.4\,GHz, and 3\,GHz, and for 0.3\,GHz for $t>5$\,yr, we found that $\nu_{\rm obs}>\nu_{\rm a}>\nu_{\rm m}$. In this case we computed the expected flux using}
\begin{equation}
F_{\rm obs}(t) = F_{\rm m}(t)\cdot \left(\frac{\nu_{obs}}{\nu_{m}(t)}\right)^{{-\frac{p-1}{2}}},
\end{equation}
where
\begin{equation}
F_{m}(t)= 500\,\mu\mbox{Jy}\,R(t)_{17}^3 n_{-1}^{3/2} \beta(t)^{1} d_{27}^{-2}
\end{equation}
\add{For $\nu_{\rm obs}=0.3$\,GHz and $t<5$\,yr we found $\nu_{\rm a}>\nu_{\rm obs}>\nu_{\rm m}$. In this case we approximated the radio flux to be}
\begin{equation}
F_{\rm obs}(t) = F_{\rm m}(t)\cdot \left(\frac{\nu_{a}(t)}{\nu_{m}(t)}\right)^{{-\frac{p-1}{2}}}
\cdot \left(\frac{\nu_{obs}}{\nu_{a}(t)}\right)^{{-\frac{5}{2}}}.
\end{equation}
\add{This approximation assumes that radio emission before its peak, which we confirmed to be the case here.}

\subsection{Model fitting}

We carried out a model fitting in which we identified the parameters of the  radio emission model described above that best matched the observed data. We found the best match by minimizing the least-squares error of the model prediction vs. the observations using gradient descent. Our best fit parameters are: ejecta mass $M_{0}=0.005$\,M$_\odot$, characteristic velocity $v_{0}\approx0.3c$, circum-merger density $n=5\,$cm$^{-3}$, and a merger time in 1993. Our model also fit the fraction \citep{2013MNRAS.430.2121P} of the kinetic energy in the shocked gas carried by electrons ($\epsilon_{\rm e}$) and magnetic fields ($\epsilon_{\rm B}$), for which we found $\epsilon_{\rm B}\approx\epsilon_{\rm e}\approx0.2$. 

\begin{figure}
\includegraphics[width=0.5\textwidth]{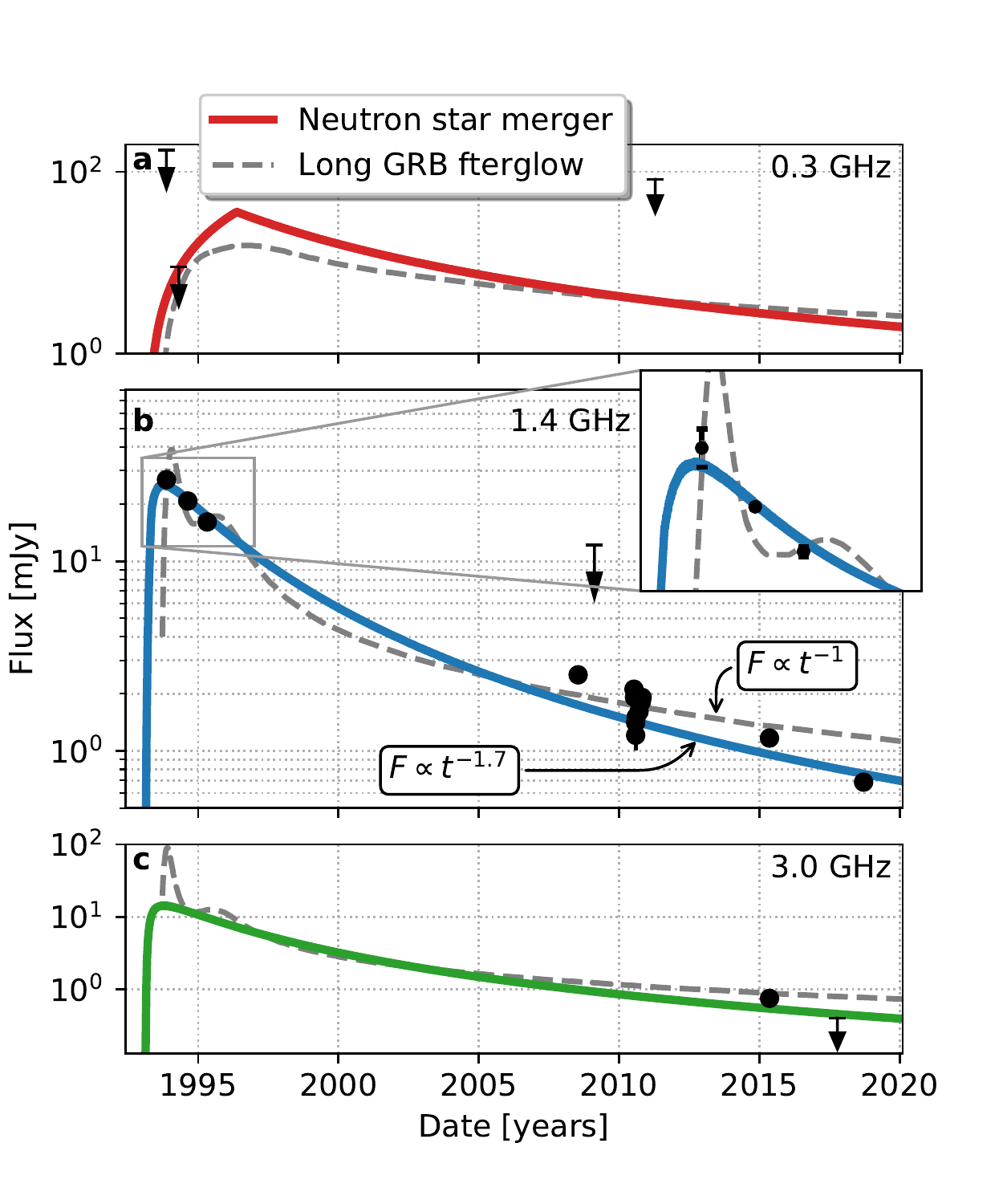}
\centering
\caption{Observational data for FIRST J1419+3940 and best fit radio light curves. The solid lines show our best fit for a neutron star merger ejecta model for 0.3\,GHz (a), 1.4\,GHz (b) and 3\,GHz (c) radio frequencies. For comparison we show the best fit model of \cite{law2018discovery} who assumed that the source is an off-axis GRB afterglow (dashed line). The inset shows the early emission period at 1.4\,GHz, indicating that radio emission steadily decreased with time, as expected from a neutron star merger scenario. The double-peaked fit for off-axis GRB afterglows less-adequately explains the data. The late temporal decay of the radio flux is characteristic of our expectations from the neutron star merger ($F\propto t^{-1.7}$), compared to the shallower decay from afterglows ($F\propto t^{-1}$).}
\label{fig:radio}
\end{figure}

The obtained radio light curves in comparison to observations are shown in Fig. \ref{fig:radio}. We used three different radio bands, 0.3\,GHz, 1.4\,GHz and 3\,GHz, where observations were available. Observations with close but different frequencies were scaled to these values. We see that the fit closely follows the data for all three radio frequencies and all times.

\subsection{Information in the Host Galaxy Properties}

We examined what information the host galaxy of FIRST J1419+3940 carries of its origin. FIRST J1419+3940 was detected in a dwarf galaxy, SDSS J141918.81+394035.8 \citep{2018ApJS..235...42A}. The galaxy has an estimated stellar mass of $\sim2\times10^7$\,M$_{\odot}$, and a star formation rate of $\sim 0.1$\,M$_{\odot}$yr$^{-1}$. Neutron star mergers are typically expected to occur in more massive galaxies as the merger rate is primarily correlated to stellar mass due to the often long delay between star formation and binary merger \citep{2014ARA&A..52...43B,2019arXiv191004890A}. \add{However, detectability through a radio flare also means that the merger needed to reside in a relatively dense circum-merger medium, introducing selection effects that are difficult to account for.} In addition, the original identification of the source is biased as \cite{2017ApJ...846...44O} searched for low-mass, star-forming galaxies \add{although this strategy by itself was likely not a major contributor to selection effects}. Therefore, here we simply assessed whether similar galaxies can be sources of neutron star mergers. 

We computed the expected merger rate in the IllustrisTNG cosmological hydrodynamical simulation \citep{2018MNRAS.475..676S, 2019ComAC...6....2N}. IllustrisTNG uses a comprehensive galaxy formation model to make self-consistent predictions for galactic stellar mass buildup within galaxies. The model itself is tuned to match a number of observational constraints including the galaxy stellar mass function and cosmic star formation rate density \citep{2013MNRAS.436.3031V, 2014MNRAS.438.1985T}. One result from the IllustrisTNG simulation is star formation rate histories for all simulated systems. Since the overall stellar mass buildup in these simulations matches stellar mass functions across a broad redshift range, the star formation histories are likely reflective of the real Universe.

We adopted the star-formation rate histories from the IllustrisTNG simulation and convolved them with an assumed delay time distribution to calculate neutron star merger rates. For our merger delay time distribution, we used a power law form with a cut-off time before which no neutron star mergers occur. We adopted the power law from \cite{2018MNRAS.477.1206N} with an exponential slope of $\gamma=-1.12$ taken to match the observed SNIa delay time distribution \citep{2012MNRAS.426.3282M}. 
We used a cut-off time of 100\,Myr as a fiducial value \citep{2019ApJ...878...12S}.

We evaluated the neutron star merger rate across the entire simulated volume, and used this to determine the fraction of neutron star merger events that occur within galaxies of interest for this paper.

Considering galaxies with stellar mass $<10^{8}$\,M$_\odot$, we found that $\sim 1\%$ of neutron star mergers are expected to occur in such galaxies. Therefore, while most neutron star mergers will occur in massive galaxies, the observed host galaxy could have been the host of a neutron star merger. 

\add{We note that other caveats remain, however. First, as the fiducial neutron star merger delay time of 100\,Myr is longer than the typical duration of star formation epochs in dwarf galaxies, the host galaxy's unusually high star formation rate is likely unrelated to the binary's formation. Second, the transient's spatial offset from the galactic center is smaller than typically expected from neutron star mergers. Both of these properties are more typical for the case of long GRBs. Nonetheless, selection effects may be important here too. High star formation may correspond to high interstellar medium densities, necessary for a bright, observable radio flare. Similarly, a larger natal kick that push the binary neutron star into a sparser environment, diminishing its observability.}

\subsection{\addd{The size of the radio source}}

\addd{Using the European VLBI Network, \cite{marcote2019resolving} measured the size of the radio source on September 18, 2018 to be about $1.6\pm0.3\pm0.2$\,pc, where the first error bar is the statistical uncertainty, and the second is the systematic uncertainty (derived here from their Table 1). Our best fit model gives a source size of 1.2\,pc on the same date. Therefore, we find a neutron star merger explanation to be consistent with the observed radio source size in 2018.}

\section{Merger vs long GRB Origin} \label{sec:comparison}

\subsection{Difficulty with GRB afterglow model}

\add{Both neutron star merger radio flare and long-GRB afterglow models are broadly consistent with the observed data (see Fig. \ref{fig:radio}), therefore the models' goodness of fit is insufficient to rule out one of them. However, we identified two features of the long-GRB afterglow light curve that are in tension with the data.
\begin{enumerate}[leftmargin=*]
\item {\bf Double-peak structure of early light curve.} \cite{law2018discovery} found that the best fit afterglow light curve assumes an off-axis GRB, i.e. that the GRB jet initially pointed away from Earth. Such off-axis afterglow light curves have a characteristic early double peak structures (see inset in Fig. \ref{fig:radio}). The two peaks are due to the two GRB jets become "visible" in radio at different times. The observed early radio data of J1419+3940 shows no such structure, but instead shows a gradual decay. Such a decay could be observed if the double peaked structured is observed such that the three observations are carried out serendipitously at the right "phases" of the two peaks, however, such coincidence is unlikely. A more likely explanation is that the radio emission was indeed gradually decaying during this time. \newline
Gradual decay is still consistent with an on-axis GRB afterglow. In this scenario the GRB jet is pointing towards or away from Earth from the beginning, therefore there will be no peaks in the radio light curve. However, while an on-axis GRB afterglow could explain the gradually decaying radio data observed at 1.4\,GHz, it wold predict a high initial flux that is inconsistent with the early upper limit observed at 0.3\,GHz. Therefore, an on-axis GRB afterglow can be ruled out.
\item {\bf Late time temporal decay.} The long GRB afterglow radio light curve is expected to flatten after about a year, when the bulk of the shock-accelerated electrons in the outflow become non-relativistic \citep{2013ApJ...778..107S,law2018discovery}. This results in a {\it deep Newtonian} temporal decay of the radio flux $F\propto t^{-1}$. In contrast, we found that for the neutron star merger case the bulk of shock-accelerated electrons remains ultra-relativistic until about 40\,yr after the merger. \addd{The velocity of the shock front is non-relativistic for the whole period.} This corresponds to a temporal decay of the flux $F\propto t^{-1.7}$. This latter decay seem to be the case for the observed flux of FIRST J1419+3940. \newline
Near-future follow-up observations will further differentiate these two decay types. Given the expected differences in the temporal decay of the radio flux for the neutron star and afterglow models, the predicted flux difference for an observation during the summer of 2021 is around 500\,$\mu$J for the radio frequencies considered here, which is beyond the uncertainty expected from the power-law decay fit on the 2009-2018 observations. Therefore, we found that one additional observation in the summer of 2021 with VLA can substantially constrain the deep-Newtonian explanation of the late radio light curve.
\end{enumerate}}

\subsection{A priori detection probabilities}

As an alternative probe of the source's possible origin, we computed the a priori detection probability of a neutron star merger with the strategy that uncovered FIRST J1419+3940. \add{We then compared this number to the a priori probability of a long GRB afterglow origin.}  

\add{We carried out Monte Carlo simulations in which we randomly placed neutron star mergers in space and time and checked whether the same detection strategy that was used to identify FIRST J1419+3940 would identify them.}

\add{For each neutron star merger we randomly drew (i) ejecta mass within $[10^{-2}M_\odot,10^{-1}M_\odot]$ and ejecta velocity within $0.1$c\,$-$\,$0.2$c using independent uniform distributions; (ii) circum-merger density from the distribution of reconstructed densities for short GRBs (see Fig. top left in \citealt{2015ApJ...815..102F}); (iii) location drawn from a uniform volumetric distribution within $d_{\rm max}=108$\,Mpc, which is the maximum distance in the analysis of \cite{2017ApJ...846...44O}; (iv) time of the merger within the 50\,yr period prior to the time of observation. 
Using these simulations we determined the average duration $\langle \Delta t\rangle=1.1$\,yr for which a neutron star radio flare's flux is $>4$\,mJy at 1.4\,GHz, which is the threshold of the FIRST survey (see \citealt{law2018discovery}).} 

We adopted a neutron star merger rate of $\mathcal{R}_{\rm NS}=900^{+2940}_{-790}$\,Gpc$^{-3}$yr$^{-1}$ at $\sim90\%$ confidence level (we obtained the expected value by averaging the four different analyses discussed in Section VII.C of \citealt{2019PhRvX...9c1040A}).  \add{We note here that the higher end of this range ($\lesssim10^3$\,Gpc$^{-3}$yr$^{-1}$) have been ruled out by radio transient searches \citep{2006ApJ...639..331G,2002ApJ...576..923L}.} We additionally took into account that the FIRST survey covered about $f_{\rm FIRST}\sim25\%$ fraction of the sky, that the Karl G. Jansky Very Large Array sky survey (VLASS) used by \cite{2017ApJ...846...44O} covered about $f_{\rm VLASS}\sim50\%$ fraction of the sky, and that the galaxy sample used in the study by \cite{2017ApJ...846...44O} had a completeness of $f_{\rm galaxy}\sim30\%$ out to the covered $d_{\rm max}=108$\,Mpc. With these factors the expected number of neutron star merger radio flares detected by the survey is
\begin{equation}
\mathcal{N}_{\rm NS} = \frac{4}{3}\pi d_{\rm max}^{3}\,\mathcal{R}_{\rm NS}\,\langle \Delta t\rangle\, f_{\rm FIRST}\,f_{\rm VLASS}\, f_{\rm galaxy},
\label{eq:rate}
\end{equation}
which gives us $\mathcal{N}_{\rm NS}=0.2^{+0.64}_{-0.17}$. \add{Nevertheless, we note that the model parameters above, such as the neutron star's environment, ejecta mass and and ejecta speed are poorly constrained at present. Our uncertainty of $\mathcal{N}_{\rm NS}$ is therefore larger than the statistical uncertainty quoted here.}

For comparison we adopted here the relevant estimate of the long GRB afterglow detection rate here from \cite{law2018discovery}. \add{They estimated $\langle \Delta t\rangle=2000$\,days to be the observed duration of FIRST J1419+3940 instead of our Monte Carlo simulation, and used a different source rate of $\mathcal{R}_{\rm GRB}\sim60$\,Gpc$^{-3}$yr$^{-1}$ (see also \citealt{2010MNRAS.406.1944W,2016ApJ...818...18G}), but otherwise carried out the same computation as we describe above.} They found $\mathcal{N}_{\rm GRB}=0.06$, which is likely significantly lower than the a priori neutron star merger probability.

\section{\add{Will similar events be detected in the near-future?}} \label{sec:futureobs}

Ongoing radio surveys have superior sensitivity and sky coverage compared to the FIRST survey \citep{2019arXiv190701981L}. Consequently, our results indicate that additional neutron star mergers may already be present in survey data and could be recovered in the near future. To quantify this possibility, we estimated the expected number of radio flares from neutron star mergers in the Karl G. Jansky Very Large Array Sky Survey (VLASS; \citealt{2019arXiv190701981L}), a highly sensitive survey with its first epoch completed. 

We carried out a similar Monte Carlo simulation of neutron star mergers as described above for the case of epoch 1 of the VLASS survey, which has already been completed. The detection threshold for this survey is $120\,\mu$Jy (adopted from VLASS; \citealt{2019arXiv190701981L}). We considered a maximum source distance of $d_{\rm max}=0.5$\,Gpc, assumed that a complete galaxy catalog is available or can be completed \citep{2015ApJ...801L...1B} ($f_{\rm galaxy}=1$), and took the survey's sky coverage to be $f_{\rm VLASS}=0.8$ \citep{2019arXiv190701981L}. With these parameters we obtained $\langle \Delta t\rangle=0.3$\,yr. Considering a neutron star merger rate of $\mathcal{N}_{\rm NS}=900^{+2940}_{-790}$, we found that $110^{+350}_{-96}$ neutron star mergers are expected to have been detected by VLASS during its first epoch. 

This detection, however, is not sufficient by itself to establish the origin of these radio sources. Therefore, we extended our Monte Carlo simulations to include a second observing time for all neutron star mergers and computed the number of events that can be established as transient radio sources using the VLASS epoch 1 observation and a second observation. Taking the second observation to be VLASS epoch 2, which was taken to occur 32 months after epoch 1 and have the same sensitivity of $120\,\mu$Jy, we found that only $2.5^{+8.2}_{-2.3}$ neutron star mergers will appear as transients, making this strategy risky. Therefore, we considered a targeted VLA follow-up of radio sources, with observations taking place during the summer of 2021, and assuming a sensitivity of 3\,$\mu$Jy \citep{2019MNRAS.485.4150B}. We found that such a targeted follow-up will be able to establish $17^{+56}_{-15}$ radio signals to have originated from neutron star mergers. 

\add{This number is higher than the estimate of \cite{2015ApJ...806..224M} who found that the VLASS survey is unlikely to detect neutron star merger radio flares. This result differs from our calculation by fixing the circum-merger medium density to 0.1\,cm$^{-3}$ and the ejecta mass to $10^{-2}$\,M$_\odot$. Both of these numbers are exceeded in our model by a fraction of the mergers, producing brighter flares.}

\section{Conclusion} \label{sec:conclusion}

\add{We investigated the origin of J1419+3940 by comparing expectations from a neutron star merger radio flare and a long GRB afterglow. Our conclusions are the following:
\begin{enumerate}[leftmargin=*]
\itemsep0em
\item Both radio flare and afterglow light curves are able to fit the observational data.
\item Afterglow light curves have difficulty explaining the steady early decay of the radio flux. Off-axis afterglows have a double peaked structure that need to be serendipitously sampled to observe a steady decay. On-axis afterglows are ruled out by an early 0.3\,GHz upper limit.
\item Towards the late observations, the afterglow model light curve entered a deep-Newtonian state with a shallow decay, while the neutron star merger light curve is still expected to decay faster. This difference will provide strong observable deviations between the models in the next 1-2 years.
\item The host galaxy of the source, a highly starforming dwarf galaxy is typical for long GRBs while uncommon for neutron star mergers. The small angular offset from the galactic center is also more typical for long GRB afterglows. Nonetheless, neutron stars could be produced in such galaxies, and selection effects due to the observability of the radio emission somewhat increase the probability of a neutron star origin.
\item Existing radio surveys should already include potentially dozens of neutron star merger radio flares, making the search for such sources timely.
\end{enumerate}}

Observing radio flares will deliver a wealth of information about the origin and physics of neutron star mergers. Radio localization can identify the host galaxy and the merger's position within it, deciphering the age of the stellar population, the formation channel of the binary, or displacement due to natal kicks in supernova explosions \citep{2017LRR....20....3M}. \add{If J1419+3940 was produced by a neutron star merger, its distance from the galactic center suggests that the neutron stars were in an isolated binary system. The location of J1419+3940 within a dwarf galaxy is \addd{consistent with} a low natal kick velocity.}

Additionally, recovering the properties of the ejecta mass and velocity enables us to constrain the neutron star's masses, the properties of matter at supranuclear densities---the so-called equation of state, and the neutron-star origin of heavy elements in the Universe. \add{Assuming a neutron star merger origin}, we find that the ejecta properties of J1419+3940 are typical for dynamical ejecta in the merger of two with neutron stars with roughly equal masses \citep{2018ApJ...869..130R,2013ApJ...773...78B}. For comparison, the first neutron star merger detected through gravitational waves---GW170817 \citep{2017PhRvL.119p1101A}, had a higher dynamical ejecta mass of $\gtrsim2\times10^{-2}$\,M$_\odot$, suggesting unequal neutron star masses \citep{2019EPJA...55..203S}. The relative masses for the other neutron star merger discovered through gravitational waves, GW190425 \citep{2020ApJ...892L...3A}, is also unclear. Equal-mass binaries are more common based on detected systems in the Milky Way \citep{2019Univ....5..159L}, therefore J1419+3940 may be more representative of the binary population than GW170817. 

The authors thank Andrew MacFadyen, Szabolcs Marka, Brian Metzger, and Lorenzo Sironi for useful discussions, and Casey Law for bringing the observations to the authors' attention. The authors acknowledge the support of the National Science Foundation under grant PHY-1911796, the Alfred P. Sloan Foundation, and the University of Florida.

\bibliography{Refs}
\end{document}